\documentclass[epj,referee]{svjour}
\usepackage{graphics}
\usepackage{amsmath}
\usepackage{amssymb}
\usepackage{times}
\usepackage{epsf}
\usepackage{graphicx}
\usepackage{dcolumn}
\usepackage{bm}
\usepackage{CJK}
\begin{document}
\title{Manipulating local heat flux with different patterns}
\author{N.Q. Zhu\inst{}, X.Y. Shen\inst{}, and J.P. Huang\inst{}
\thanks{\emph{email:} jphuang@fudan.edu.cn}
}                     
%
%
\institute{Department of Physics, State Key Laboratory of Surface Physics,
and Key Laboratory of Micro and Nano Photonic Structures (Ministry of Education),
Fudan University, Shanghai 200433, China}
%
%
\abstract{
 Since the thermal conduction equation has form
 invariance under coordinate transformation, one can design thermal metamaterials
 with novel functions by tailoring materials' thermal conductivities.  In this work, we  establish a different transformation theory, and
 propose a layered device with anisotropic thermal conductivities.  The device  is
 able to convert heat flux from parallel patterns into non-parallel patterns
  and vice versa. In the mean time, the heat flux pattern outside the device keeps undisturbed as if this device is absent.
  We perform finite-element simulations to confirm the converting behavior.
  This work paves a different way to manipulate the flow of heat at will.
} 
\maketitle
\section{Introduction}
\label{sec:1}
%
%
In 2008, Fan {\it et al.}~\cite{Fanapl2008} started to adopt the coordinate transformation approach to propose a class of thermal metamaterials  working as thermal cloaking, which makes heat flow  around an ``invisible'' region and eventually returns to its original pathway as if the ``invisible region'' does not exist. This cloaking originates from the fact that the thermal conduction equation remains form-invariant under coordinate transformation, and it helps to pave a new way to  steer heat flux. As a result, much attention has been paid by  theorists and experimentalists~\cite{Lijap2010,Narprl2012,Schprl2013,Hanprl2014,Xuprl2013,Heapl2013,Hinepjap2013}, and there come out a lot of thermal metamaterials with novel thermal properties beyond cloaking~\cite{Xuprl2013,Shenijht2014,Hansp2013,jiaapl2008,Gaoepl2013}, such as concentrators (which are used to enhance the temperature gradient in a specific region)~\cite{Narprl2012,Gueoe2012}, inverters (which allow heat to apparently flow from a colder to a warmer region without violating the second law of thermodynamics)~\cite{Fanapl2008,Gueoe2013}, rotators (which can rotate the flow of heat as if the heat comes from a different direction)~\cite{Narprl2012,Gueoe2013}, and camouflage~\cite{HanAM14}.

Controlling the flow of heat enlightens us to design an illusion device for thermal conduction. As known to us, by using some special approaches one can replace object A's scattering pattern by object B's, which is the basic of optical illusion~\cite{Laiprl2009}. This concept has been extended to other fields (say, acoustics~\cite{RenAMM14}) where wave equations dominate. Obviously, an illusion of thermal conduction not only offers a different way for controlling heat conduction but also has extensive applications in misleading the detectors of temperature distribution signatures. Just as our recent work shows, a thermal illusion device based on the thermal conduction equation can be created by adopting a complementary layer~\cite{Chen}. However, the thermal conductivity of the complementary layer should be negative. For complying with the second law of thermodynamics, one must apply external work on the system~\cite{Shenijht2014,Gaoepl2013}. Inspired by the former work~\cite{Chen}, in order to build thermal illusion devices without using negative-conductivity materials, here we attempt to propose a new kind of thermal illusion device. To this end, we shall change the flow direction of local heat flux inside the device whereas the pattern of heat flux outside the device keeps unchanged (as if the device does not exist). That is, a phenomenon of thermal illusion is created through the thermal illusion device. In order to check whether the device works or not, we shall resort to  two-dimensional finite-element simulations.




\section{Method}
\label{sec:2}
\subsection{Coordinate Transformation Approach}
\label{sec:3}

Considering a typical thermal conduction process,  heat flux is proportional to a temperature gradient. The thermal conduction equation can then be written as
\begin{equation}
      \rho C\frac{{\partial T}}{{\partial t}} + \nabla\cdot \left( { - \kappa \nabla T} \right) = Q,\
  \label{eq10}
\end{equation}
where $\rho$ and $C$ are the density and heat capacity respectively and $T$ represents temperature evolving with time $t$ at each point $X=(x,y)$ in the space. In equation~(\ref{eq10}), $\kappa$ is  thermal conductivity, and $Q$ is a heat source. For a steady state process, the distribution of temperature $T$ is independent of time $t$, and thus the first term in equation~(\ref{eq10}) vanishes. Throughout this work, we suppose there is no heat source,  $Q=0$. Therefore, equation~(\ref{eq10}) can be reduced to $\nabla\cdot ( - \kappa \nabla T) = 0$.

Upon a change of variable $X = (x,y)$  $\to$ $X' {\rm{ }} = {\rm{ }}\left( {x' ,y' } \right)$ described by a Jacobian transformation matrix {\bf J}, this equation, $\nabla\cdot ( - \kappa \nabla T) = 0$, takes the following form,
\begin{equation}
      \nabla\cdot \left( { -  \frac{{{\rm {\bf J}}\kappa{{\rm {\bf J}}^T}}}{{\det \left( {\rm {\bf J}} \right)}}\nabla T} \right) = 0,
  \label{eq11}
\end{equation}
where ${\rm {\bf J}}^T$ is the transposed matrix of ${\rm {\bf J}}$, and $\det({\rm {\bf J}})$ is the determinant of ${\rm {\bf J}}$.
Thus the new thermal conductivity in the transformed space $X'$ can be expressed as
 \begin{equation}
\kappa '{\rm{ = }}\frac{{{\rm {\bf J}}\kappa{{\rm {\bf J}}^T}}}{{\det \left( {\rm {\bf J}} \right)}}\equiv \left( {\begin{array}{*{20}{c}}
{{\beta_{11}}}&{{\beta_{12}}}\\
{{\beta_{21}}}&{{\beta_{22}}}
\end{array}} \right).
    \label{eq5}
\end{equation}

\subsection{Transformation~1}
\label{sec:4}

The coordinate transformation is constructed in two-dimensional Cartesian coordinate systems and schematically presented in Figure~1. Suppose the radius of the sector is $a$, the length of $OM$ is $b$ and the side length of the square is $2c$.

Similarly, we use $X = (x,y)$ to represent an arbitrary point in the original space, and  $X' {\rm{ }} = {\rm{ }}\left( {x' ,y' } \right)$ to denote the corresponding point in the transformed space. According to the geometrical relation that $r = \sqrt {{x^2} + {y^2}}~{\rm and} ~r' = \sqrt {{{x'}^2} + {{y'}^2}},\label{eq2}$ we can derive the mapping transformation in Regions~I and II,
\begin{equation}
     r'={\raise0.7ex\hbox{${ax}$} \!\mathord{\left/
     {\vphantom {{ax} b}}\right.\kern-\nulldelimiterspace}
     \!\lower0.7ex\hbox{$b$}}
     \qquad (0 \le x \le b).
     \label{eq1}
\end{equation}
Clearly, equation~(\ref{eq1}) is used to geometrically compress the triangle $OCD$ (original space) to the sector $\widehat{OAB}$ (transformed space). Accordingly the vertical lines in the original space is distorted to arcs in the transformed space. If we fill Regions~I and II  with appropriate anisotropic thermal conductivities developed by transformation norms~\cite{Fanapl2008},  parallel heat flux propagating through these regions can be converged.

Equation~(\ref{eq1}) can then be easily derived as
\begin{equation}
      \left( \begin{array}{l}
      {x'}\\
      {y'}
      \end{array} \right) = \left( \begin{array}{l}
      {\raise0.7ex\hbox{${a{x^2}}$} \!\mathord{\left/
      {\vphantom {{a{x^2}} {rb}}}\right.\kern-\nulldelimiterspace}
      \!\lower0.7ex\hbox{${rb}$}}\\
      {\raise0.7ex\hbox{${axy}$} \!\mathord{\left/
       {\vphantom {{bxy} {rb}}}\right.\kern-\nulldelimiterspace}
      \!\lower0.7ex\hbox{${rb}$}}
      \end{array} \right)
      \qquad (0 \le x \le b).
      \label{eq3}
\end{equation}
Therefore, we are able to obtain the anisotropic thermal conductivity tensor $\kappa_1'$ of the illusion device,
 \begin{equation}
{\kappa_1 ^\prime } = \left( {\begin{array}{*{20}{c}}
{1 + {\raise0.7ex\hbox{${3{y^2}}$} \!\mathord{\left/
 {\vphantom {{3{y^2}} {{r^2}}}}\right.\kern-\nulldelimiterspace}
\!\lower0.7ex\hbox{${{r^2}}$}}}&{{\raise0.7ex\hbox{${2y}$} \!\mathord{\left/
 {\vphantom {{2y} x}}\right.\kern-\nulldelimiterspace}
\!\lower0.7ex\hbox{$x$}} - {\raise0.7ex\hbox{${3xy}$} \!\mathord{\left/
 {\vphantom {{3xy} {{r^2}}}}\right.\kern-\nulldelimiterspace}
\!\lower0.7ex\hbox{${{r^2}}$}}}\\
{{\raise0.7ex\hbox{${2y}$} \!\mathord{\left/
 {\vphantom {{2y} x}}\right.\kern-\nulldelimiterspace}
\!\lower0.7ex\hbox{$x$}} - {\raise0.7ex\hbox{${3xy}$} \!\mathord{\left/
 {\vphantom {{3xy} {{r^2}}}}\right.\kern-\nulldelimiterspace}
\!\lower0.7ex\hbox{${{r^2}}$}}}&{{\raise0.7ex\hbox{${{r^2}}$} \!\mathord{\left/
 {\vphantom {{{r^2}} {{x^2}}}}\right.\kern-\nulldelimiterspace}
\!\lower0.7ex\hbox{${{x^2}}$}} - {\raise0.7ex\hbox{${3{y^2}}$} \!\mathord{\left/
 {\vphantom {{3{y^2}} {{r^2}}}}\right.\kern-\nulldelimiterspace}
\!\lower0.7ex\hbox{${{r^2}}$}}}
\end{array}} \right) .
\label{eq7}
\end{equation}
The values of the components of $\kappa_1' $  have no relation with either $a$ or $b$ and are determined only by the coordinates of the points. Note there is $\det(\kappa_1')$=1, which is positive. It should also be noted that the conductivity tensor above satisfy a symmetrical matrix form
$$\left(
  \begin{array}{cc}
    \kappa_{xx} & \kappa_{xy} \\
    \kappa_{xy} & \kappa_{yy} \\
  \end{array}
\right).$$
To illustrate this conductivity tensor is positive and provide practicable parameters for experiment, we can diagonalize it by rotating the principle axis by an appropriate angle $\alpha$. This angle can be calculated by using the relation: $$\alpha=\frac{1}{2}\arctan\frac{2\kappa_{xy}}{\kappa_{xx}-\kappa_{yy}},[-\frac{\pi}{2}<\alpha<\frac{\pi}{2}],$$ where $\kappa_{xy},\kappa_{xx},\kappa_{yy}$ are the elements of the tensor $\kappa_1'$.
As a result, we obtain the diagonalized tensor as
$$\left(
  \begin{array}{cc}
    \frac{2x^{2}+y^{2}+y(4x^{2}+y^{2})^{\frac{1}{2}}}{2x^{2}} & 0 \\
    0 & \frac{2x^{2}+y^{2}-y(4x^{2}+y^{2})^{\frac{1}{2}}}{2x^{2}}  \\
  \end{array}
\right).$$
Obviously, owing to  $2x^{2}+y^{2}>y(4x^{2}+y^{2})^{\frac{1}{2}}$,  the tensor is positive as expected.

\subsection{Transformation~2}
\label{sec:5}

Transformation~1 mentioned above deals with the steady-state thermal conduction and converges the parallel heat flux on the vertex of the triangle, namely, from  parallel patterns to non-parallel patterns. For the sake of completeness, here we want to attempt an inverse behavior, i.e., from  non-parallel patterns to parallel patterns. To proceed, we assume a temperature field associated with an extremely small ring located at the centre where  heat flow diverges from it, and we aim to design a device converting the dispersed heat flux (non-parallel patterns) into parallel patterns.

A proper approach to achieve this goal is to make the following transformation that is illustrated in Figure~2. Suppose the radius of the small circle is $a$, the side length of the square is $2b$ and the radius of the big circle is $\sqrt{2}b$. We extend the sector $\widehat{OAB}$ (original space) to the triangle $OCD$ (transformed space). The geometrical relation of the transformation means that each point in an arc located in the original space will be mapped to a vertical segment in the transformed space. As a result, when the divergent heat flux reaches the device, they will change to a parallel pattern due to the distorted coordinates.

Also, we use $X = (x,y)$ and $X' {\rm{ }} = {\rm{ }}\left( {x' ,y' } \right)$ to respectively represent an arbitrary point in the original space and the corresponding point in the transformed space. We obtain the mapping transformation in Regions~I and III,
\begin{equation}
\left( \begin{array}{l}
{{\rm{x'}}}\\
{y'}
\end{array} \right){\rm{ = }}\left( \begin{array}{l}
{\raise0.7ex\hbox{${br}$} \!\mathord{\left/
 {\vphantom {{br} a}}\right.\kern-\nulldelimiterspace}
\!\lower0.7ex\hbox{$a$}}\\
{\raise0.7ex\hbox{${bry}$} \!\mathord{\left/
 {\vphantom {{bry} {bx}}}\right.\kern-\nulldelimiterspace}
\!\lower0.7ex\hbox{${ax}$}}
\end{array} \right)\quad {\rm{(r}} \le {\rm{a)}}.
      \label{eq6}
\end{equation}

The thermal conductivity $\kappa_2'$ for Regions~I and III can be derived from equation~(\ref{eq5}),
\begin{equation}
{\kappa _2}^\prime  = \left( {\begin{array}{*{20}{c}}
{{\raise0.7ex\hbox{${{x^2}}$} \!\mathord{\left/
 {\vphantom {{{x^2}} {{r^2}}}}\right.\kern-\nulldelimiterspace}
\!\lower0.7ex\hbox{${{r^2}}$}}}&{{\raise0.7ex\hbox{${xy}$} \!\mathord{\left/
 {\vphantom {{xy} {{r^2}}}}\right.\kern-\nulldelimiterspace}
\!\lower0.7ex\hbox{${{r^2}}$}}}\\
{{\raise0.7ex\hbox{${xy}$} \!\mathord{\left/
 {\vphantom {{xy} {{r^2}}}}\right.\kern-\nulldelimiterspace}
\!\lower0.7ex\hbox{${{r^2}}$}}}&{{\raise0.7ex\hbox{${{r^2}}$} \!\mathord{\left/
 {\vphantom {{{r^2}} {{x^2}}}}\right.\kern-\nulldelimiterspace}
\!\lower0.7ex\hbox{${{x^2}}$}} + {\raise0.7ex\hbox{${{y^2}}$} \!\mathord{\left/
 {\vphantom {{{y^2}} {{r^2}}}}\right.\kern-\nulldelimiterspace}
\!\lower0.7ex\hbox{${{r^2}}$}}}
\end{array}} \right) .
\label{eq8}
\end{equation}

With the help of the rotation matrix, we can derive the thermal conductivity $\kappa_2''$ for Regions~II and IV,
\begin{equation}
{\kappa _2''} = \left( {\begin{array}{*{20}{c}}
{{\raise0.7ex\hbox{${{r^2}}$} \!\mathord{\left/
 {\vphantom {{{r^2}} {{x^2}}}}\right.\kern-\nulldelimiterspace}
\!\lower0.7ex\hbox{${{x^2}}$}} + {\raise0.7ex\hbox{${{y^2}}$} \!\mathord{\left/
 {\vphantom {{{y^2}} {{r^2}}}}\right.\kern-\nulldelimiterspace}
\!\lower0.7ex\hbox{${{r^2}}$}}}&{{\raise0.7ex\hbox{${xy}$} \!\mathord{\left/
 {\vphantom {{xy} {{r^2}}}}\right.\kern-\nulldelimiterspace}
\!\lower0.7ex\hbox{${{r^2}}$}}}\\
{{\raise0.7ex\hbox{${xy}$} \!\mathord{\left/
 {\vphantom {{xy} {{r^2}}}}\right.\kern-\nulldelimiterspace}
\!\lower0.7ex\hbox{${{r^2}}$}}}&{{\raise0.7ex\hbox{${{y^2}}$} \!\mathord{\left/
 {\vphantom {{{y^2}} {{r^2}}}}\right.\kern-\nulldelimiterspace}
\!\lower0.7ex\hbox{${{r^2}}$}}}
\end{array}} \right) .
\label{eq9}
\end{equation}

Also, the values of the components of $\kappa_2'$ and $\kappa_2''$ do not have any relation with either $a$ or $b$ and are only given  by the coordinates of the points. Incidentally, there  exist $\det(\kappa_2')=1$ and $\det(\kappa_2'')=1$. That is, the two tensors are both positive, which  can also be confirmed by using the same diagonalization method as we adopted at the end of Section 2.2 (for Transformation~1). In addition, since Transformation~2 is just a reverse operation of Transformation~1, the diagonalized conductivity tensor of Transformation~2 would be as same as what we derived for Transformation~1. The new diagonalized conductivity tensor of $\kappa_2'$ is
$$\left(
  \begin{array}{cc}
    \frac{2x^{2}+y^{2}+y(4x^{2}+y^{2})^{\frac{1}{2}}}{2x^{2}} & 0 \\
    0 & \frac{2x^{2}+y^{2}-y(4x^{2}+y^{2})^{\frac{1}{2}}}{2x^{2}}  \\
  \end{array}
\right),$$
and the diagonalized conductivity tensor of $\kappa_2''$ is
$$\left(
  \begin{array}{cc}
    \frac{2y^{2}+x^{2}+x(4y^{2}+x^{2})^{\frac{1}{2}}}{2y^{2}} & 0 \\
    0 & \frac{2y^{2}+x^{2}-x(4y^{2}+x^{2})^{\frac{1}{2}}}{2y^{2}}  \\
  \end{array}
\right).$$

\section{Results}
\label{sec:6}

We perform finite element simulations based on commercial software COMSOL Multiphysics ($http://www.comsol.com$) to check whether the device works indeed.

\subsection{Transformation~1}
\label{sec:7}

Figure~3a illustrates the temperature distribution in the original space, where the thermal conductivity $\kappa_1$ is $1$\,W/\,(m$\cdot$K). The temperature of the left boundary is set to be $273.15$\,K while the right to be $373.15$\,K. The upper and lower boundaries are set to be thermal insulation. The white arrows representing the heat flux are parallel.

The temperature distribution of Transformation~1 is shown in Figure~3b. We set the following parameters: $a=0.20$\,m, $b=0.25$\,m, and $c=0.50$\,m. Regions~I and II are filled up with $\kappa_1'$ from equation~(\ref{eq7}). The thermal conductivity of the rest space and the boundary conditions are same as what we have adopted in Figure~3a. From Figure~3b, we can see that the parallel heat flux has become converged towards the vertex of the triangle area. Moreover, the device has almost no influence on the temperature distribution outside. The pattern of arrows in the rest space are same as those in Figure~3a. For a quantitative understanding of the parameter setting, we plot Figures~4a-4c, which show the specific values of the components of the conductivity tensor $\kappa_1'$.

\subsection{Transformation~2}
\label{sec:8}

In Figure~5a, we show the temperature distribution of a small circle settled in a host medium. Both of them are filled up with isotropic thermal conductivity $\kappa_2$, which is set as $5$\,W/\,(m$\cdot$K). Additionally, we use an extremely small ring ($r_0=0.002$\,m) at the centre and the temperature of its edge is $493.15$\,K. The boundary condition of the simulation region is a constant temperature $293.15$\,K. Hence, the resulting heat flux satisfies a divergent pattern.

To show how Transformation~2 works, for Figure~5b, we set the following parameters: $a=0.50$\,m and $b=0.50$\,m. Regions~I and III are filled up with a material whose thermal conductivity tensor is $\kappa'_2$ derived from equation~(\ref{eq8}). Both the thermal conductivity of the rest space and the boundary condition are same as what we have adopted in Figure~5a. The role of Transformation~2 is shown  in Figure~5b. Within the circle area (whose boundary is indicated by a white circle), the heat flux is represented by parallel white arrows in four directions. However, outside the square, the arrows convert to divergent forms which are almost the same as those in Figure~5a. Different from Transformation~1, we succeed in dealing with the field of the extremely small ring located at the centre. The specific values of the thermal conductivity tensors $\kappa_2'$ and $\kappa_2''$ are shown in Figures~6a-6c.

\section{Conclusion}
\label{sec:9}
Since all the positive conductivity tensors we derived above are symmetrical matrices, which can be diagonalized by rotating the principle axis by an appropriate angle. The new simplified conductivity matrices can be easily used to implement the device with  effective medium theories~\cite{HuangPR06}. Thus, it should be more convenient to build such a thermal illusion device.



In a word, by adopting the coordinate transformation approach of heat conduction, we have designed a thermal illusion device composed of  materials with positive conductivity tensors. The device is able to convert the parallel pattern of heat flux into non-parallel pattern (Figure~3) and vice versa (Figure~5). Meanwhile, the heat flux (or temperature distribution) outside the device keeps the same as if the device does not exist. The two-dimensional finite-element simulations have been used to confirm the desired effects. This work not only gives an approach to control  heat flux,  but also provides a new method to create thermal illusion facilities without adopting negative thermal conductivities.
\\ \\ \\
We acknowledge the financial support by the National Natural Science Foundation of China under Grant No.~11222544, by the Fok Ying Tung Education Foundation under Grant No.~131008, by the Program for New Century Excellent Talents in University (NCET-12-0121), by the CNKBRSF under Grant No.~2011CB922004, and by the National Fund for Talent Training in Basic Science (No.~J1103204).

\clearpage
\newpage
%
%

\clearpage
\newpage

{\bf Figure captions}

\textbf{Fig.~1.}  Schematic diagram of Transformation~1. The radius of the sector is $a$, the length of $OM$ is $b$, and the side length of the square is $2c$.

\textbf{Fig.~2.}  Schematic diagram of Transformation~2. The radius of the inner (small) circle is $a$, the side length of the square is $2b$, and the radius of the outer (big) circle is $\sqrt{2}b$.

\textbf{Fig.~3.} Results of two-dimensional finite-element simulations. (a) The temperature distribution and the heat flux with $\kappa_1 =1$\,W/\,(m$\cdot$K). The parameters are set as $a=0.20$\,m, $b=0.25$\,m, and $c=0.50$\,m. The temperature of the left boundary is set to be $273.15$\,K while the right is $373.15$\,K. The upper and lower boundaries are thermal insulation. The heat flux is represented by white arrows. (b) Simulation results of Transformation~1. The parameters and the boundary conditions are the same as those in (a). Region~I is filled up with material whose thermal conductivity $\kappa_1'$ is given by equation~(\ref{eq5}). The thermal conductivity of the rest area is $1$\,W/\,(m$\cdot$K). Obviously, the heat flow represented by white arrows converges in Regions I and II.

\textbf{Fig.~4.}  Specific values of the conductivity tensor ($\kappa_1'$) for Transformation~1. The color bar represents the value of the conductivity at a certain point. (a) Illustration of  $\bf{\beta_{11}}$ in equation~(\ref{eq5}) for Transformation~1, namely, $1+3y^2/r^2$. (b) Illustration of  $\bf{\beta_{12}}$ or $\bf{\beta_{21}}$ in equation~(\ref{eq5}) for Transformation~1, i.e., $2y/x-3xy/r^2$. Because the tensor matrix $\kappa_1'$ is symmetric, $\bf{\beta_{12}}$ and $\bf{\beta_{21}}$ are equal. (c) Illustration of $\bf{\beta_{22}}$ in equation~(\ref{eq5}) for Transformation~1, i.e., $r^2/x^2 - 3y^2/r^2$.

\textbf{Fig.~5.} Results of two-dimensional finite-element simulations. (a) The temperature distribution for a material with conductivity $\kappa_2=5$\,W/\,(m$\cdot$K). The parameters are set as $a=0.50$\,m and $b=0.50$\,m. An extremely small ring  is located at the centre with temperature ($T$) at its edge, $T=493.15$\,K; the boundary condition   is set as $T=293.15$\,K. The white arrows are diverging from the centre. (b) Simulation result of Transformation~2. Both the parameters and the boundary condition are the same as those in (a). The conductivities of  materials filled in the circle (whose boundary is a white circle) are $\kappa_2'$ and $\kappa_2''$, given by equation~(\ref{eq8}) and equation~(\ref{eq9}), and the  thermal conductivity in the rest space is  $\kappa_2 =5$\,W/\,(m$\cdot$K). The conditions of the extremely small ring and the boundary are same as those in (a). Within the region of the white circle, the heat flux represented by white arrows is distorted to a parallel pattern.

\textbf{Fig.~6.} Specific values of the conductivity tensors  ($\kappa_2'$ and $\kappa_2''$) for Transformation~2.  (a) Illustration of  $\bf{\beta_{11}}$ in equation~(\ref{eq5}) for Transformation~2, namely, $x^2/r^2$ (for Regions I and III) and $r^2/x^2 + y^2/r^2$ (for Regions II and IV). (b) Illustration of $\bf{\beta_{12}}$ and $\bf{\beta_{21}}$ in equation~(\ref{eq5}) for Transformation~2, i.e., $xy/r^2$ (for Regions I-IV). Because the tensor matrix $\kappa_2'$ and $\kappa_2''$ are symmetric, $\bf{\beta_{12}}$ and $\bf{\beta_{21}}$ are equal. (c) Illustration of $\bf{\beta_{22}}$ in equation~(\ref{eq5}) for Transformation~2, namely, $r^2/x^2 + y^2/r^2$ (for Regions I and III) and $y^2/r^2$ (for Regions II and IV).  Note that (c) can be obtained by rotating (a) for $90$ degrees.


\clearpage
\newpage

%
\begin{figure}
\resizebox{1.5\columnwidth}{!}{%
  \includegraphics{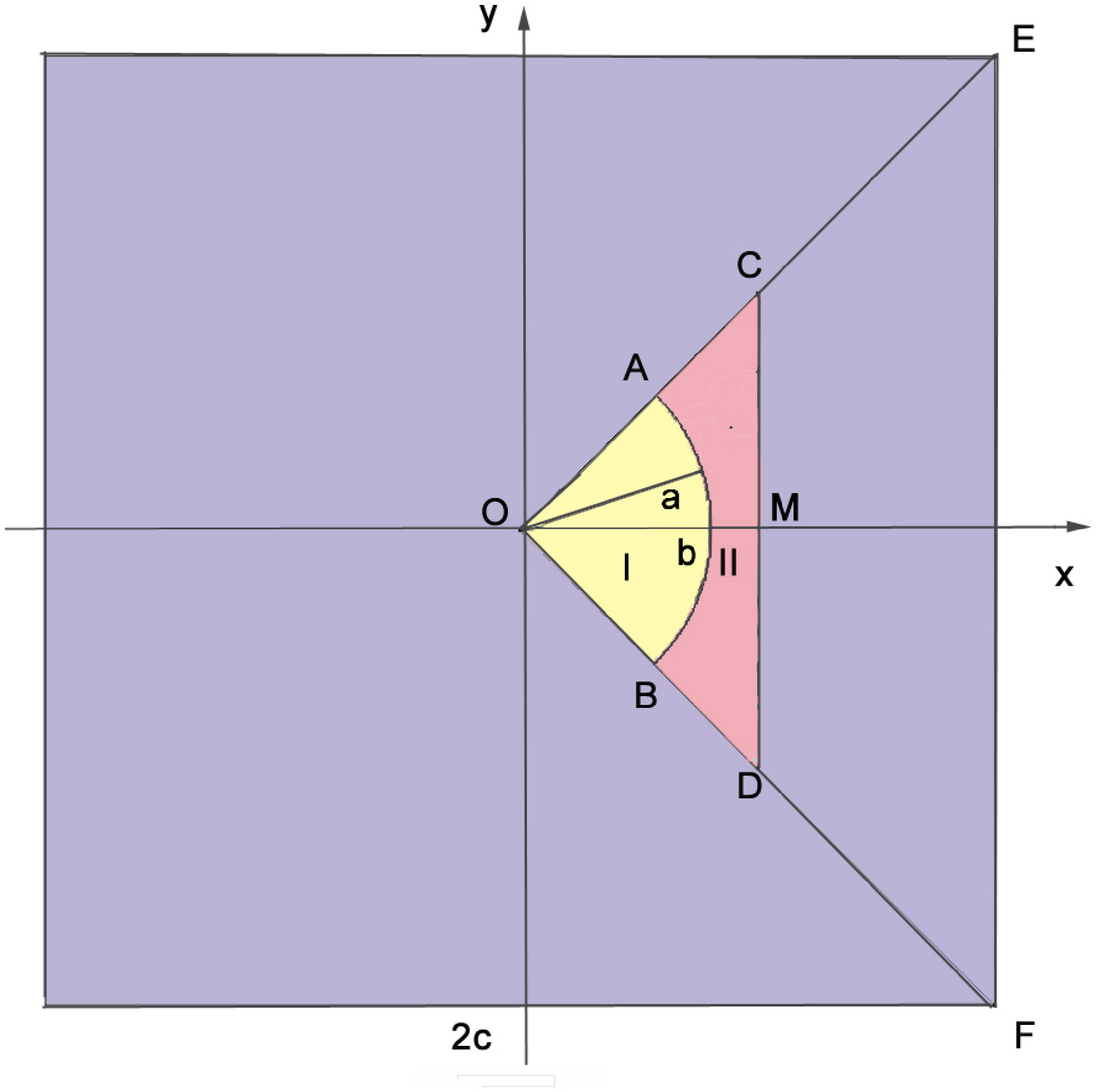}
}
\begin{center}
\caption{/Zhu, Shen, and Huang}
\label{fig:1}
\end{center}
\end{figure} 
\clearpage
\newpage
\begin{figure}
\resizebox{1.5\columnwidth}{!}{%
  \includegraphics{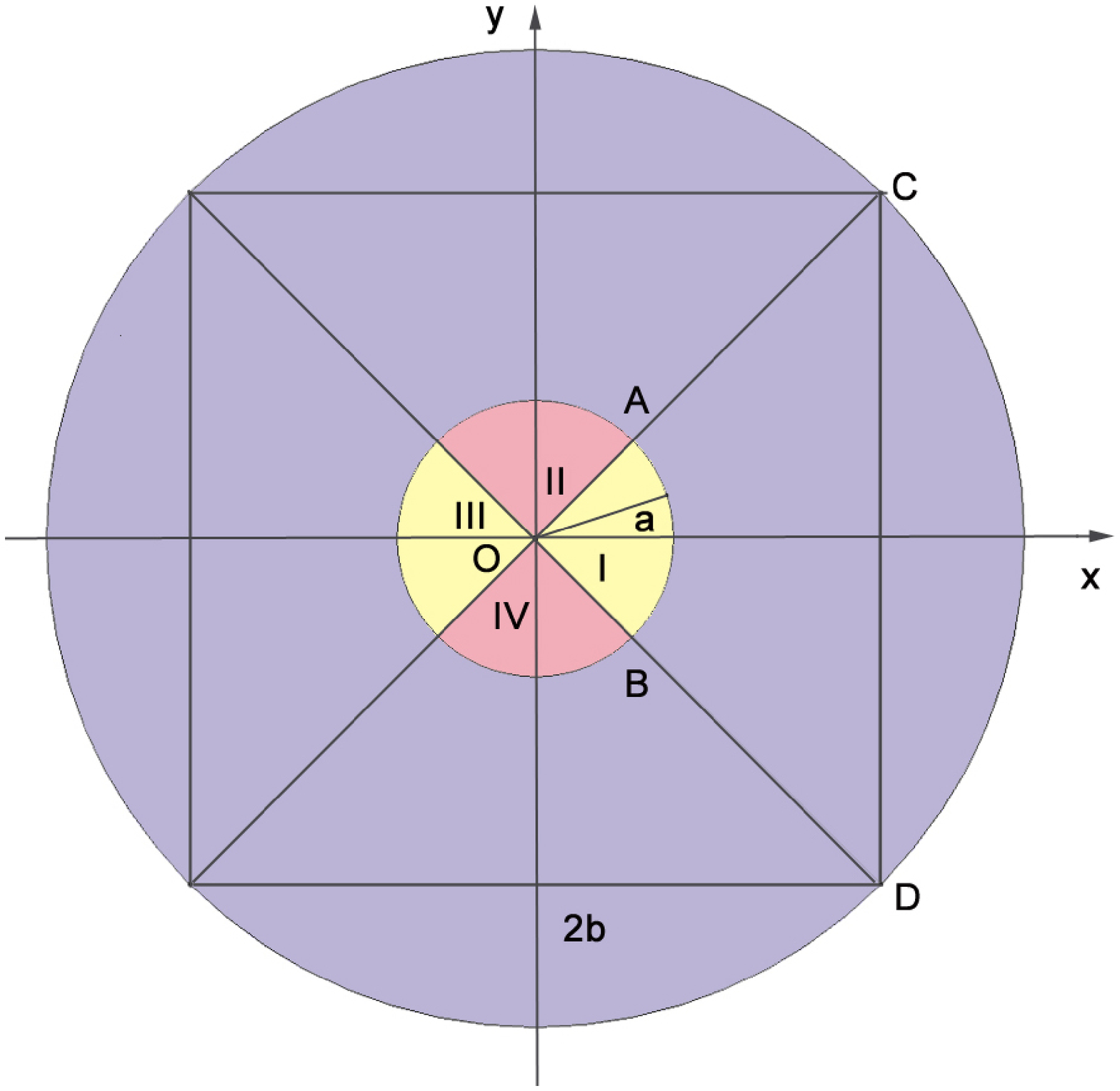}
}
\begin{center}
\caption{/Zhu, Shen, and Huang}
\label{fig:2}
\end{center}
\end{figure} 

\clearpage
\newpage
\begin{figure}
\resizebox{1.5\columnwidth}{!}{%
  \includegraphics{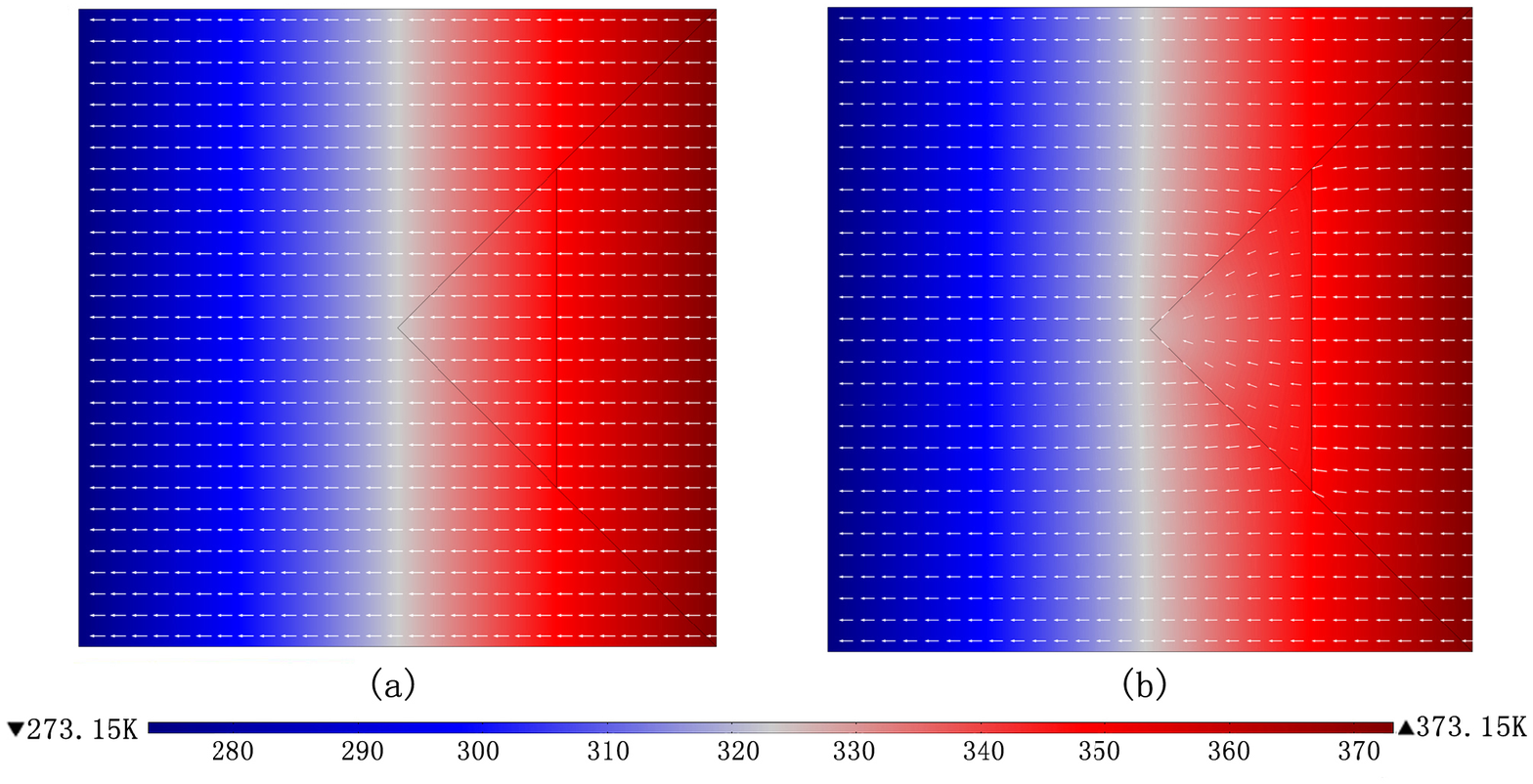}
}
\begin{center}
\caption{/Zhu, Shen, and Huang}
\label{fig:3}
\end{center}
\end{figure} 

\clearpage
\newpage
\begin{figure}
\resizebox{1.5\columnwidth}{!}{%
  \includegraphics{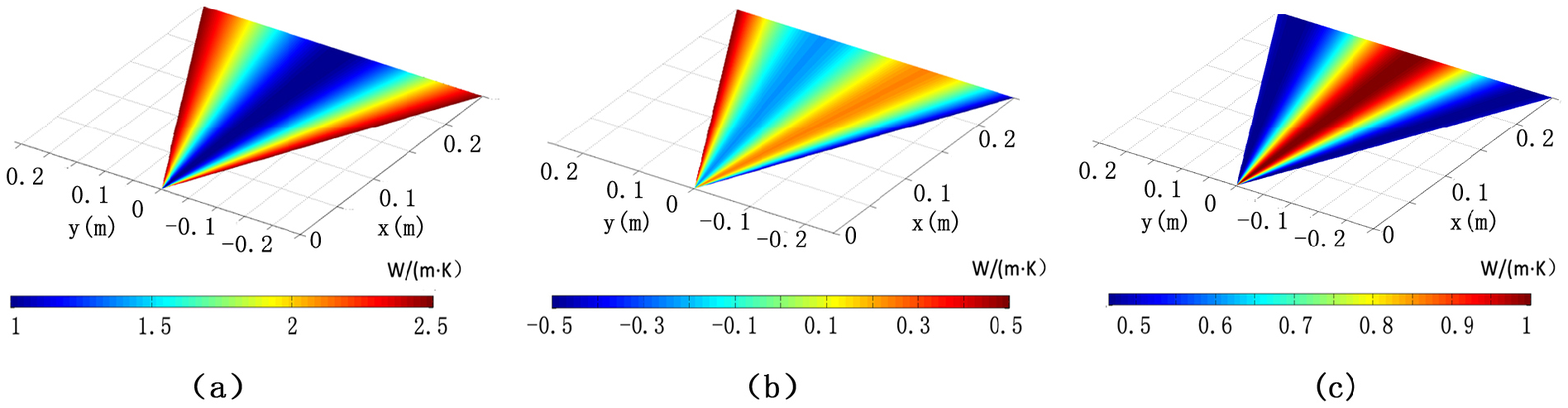}
}
\begin{center}
\caption{/Zhu, Shen, and Huang}
\label{fig:4}
\end{center}
\end{figure} 

\clearpage
\newpage
\begin{figure}
\resizebox{1.5\columnwidth}{!}{%
  \includegraphics{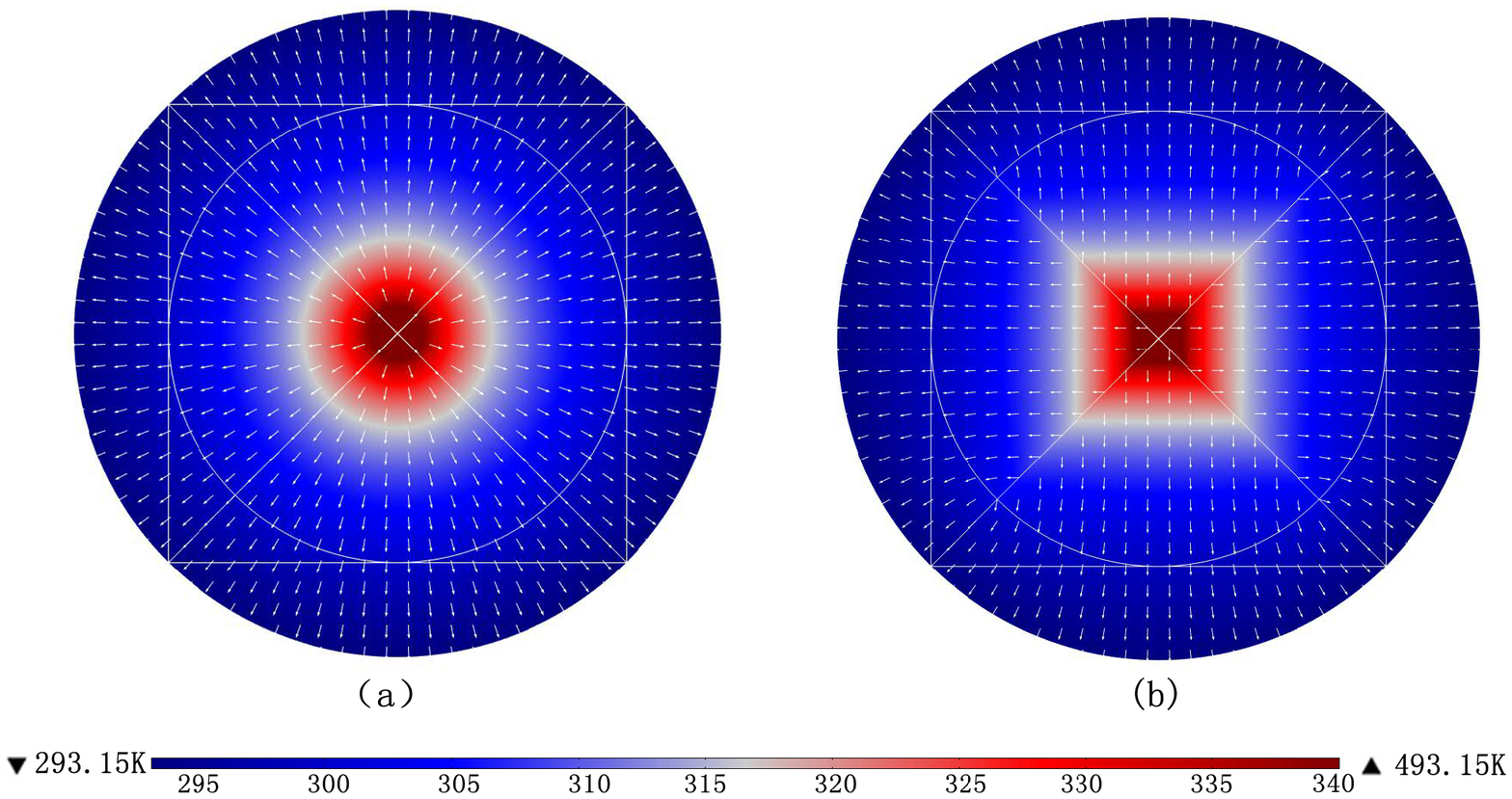}
}
\begin{center}
\caption{/Zhu, Shen, and Huang}
\label{fig:5}
\end{center}
\end{figure} 

\clearpage
\newpage
\begin{figure}
\resizebox{1.5\columnwidth}{!}{%
  \includegraphics{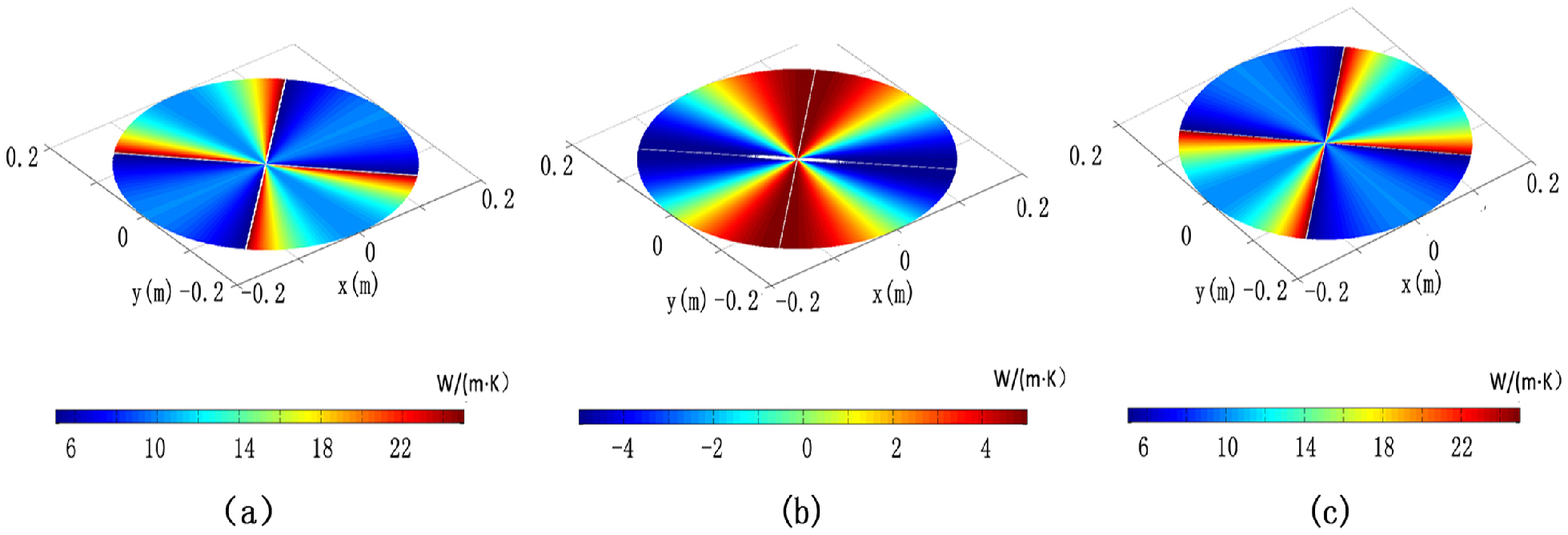}
}
\begin{center}
\caption{/Zhu, Shen, and Huang}
\label{fig:6}
\end{center}
\end{figure} 

\end{document}